\documentstyle[epsfig,12pt]{elsart} 
\begin{document}
\newcommand{\kp}{K^+}
\newcommand{\gk}{\vec{\gamma}\vec{k}}
\newcommand{\gE}{\gamma_0 E_k}
\newcommand{\ppl}{\vec{p}}
\newcommand{\bcm}{\vec{b}^{\star}}
\newcommand{\becm}{\vec{\beta}^{\star}}
\newcommand{\bepl}{\vec{\beta}}
\newcommand{\rcm}{\vec{r}^{\star}}
\newcommand{\rpl}{\vec{r}}
\newcommand{\A}{{$\mathcal A$}}
\newcommand{\wpk}{ \omega_{p-k}}
\newcommand{\Journal}[4]{ #1 {\bf #2} (#4) #3}
\newcommand{\NPA}{Nucl.\ Phys.\ A}
\newcommand{\PLB}{Phys.\ Lett.\ B}
\newcommand{\PRC}{Phys.\ Rev.\ C}
\newcommand{\ZPC}{Z.\ Phys.\ C}
\newcommand{\be}{\begin{equation}}
\newcommand{\ee}{\end{equation}}
\begin{frontmatter}

\title{In-Medium Effects on Particle Production in
Heavy Ion Collisions}

\author[auth]{V. Prassa}
\author[catania]{G.Ferini}
\author[giessen]{T. Gaitanos}
\author[muenchen]{H.H. Wolter}
\author[auth]{G.A. Lalazissis}
\author[catania]{M. Di Toro}

\address[auth]{Department of Theoretical Physics,
Aristotle University of Thessaloniki, Thessaloniki Gr-54124,Greece}
\address[catania]{Laboratori Nazionali del Sud INFN, I-95123 Catania, Italy}
\address[giessen]{Institut f\"{u}r Theoretische Physik, 
Justus-Liebig-Universit\"{a}t Giessen, 
D-35392 Giessen, Germany}
\address[muenchen]{Sektion Physik, Universit\"at M\"unchen, 
D-85748 Garching, Germany}  
\address{email: Theodoros.Gaitanos@theo.physik.uni-giessen.de}
\begin{abstract}
The effect of possible in-medium modifications of nucleon-nucleon ($NN$)
cross sections  on particle production is
investigated in heavy ion collisions ($HIC$)
at intermediate energies. In particular, using a fully covariant relativistic 
transport approach, we see that the density dependence of the
{\it inelastic} cross sections appreciably affects the pion and kaon
yields and their rapidity distributions. However, the
$(\pi^{-}/\pi^{+})$- and $(K^{0}/K^{+})$-ratios depend only moderately
on the in-medium behavior of the inelastic cross sections. This is 
particularly true for kaon yield ratios, since kaons are more uniformly
produced in high density regions. Kaon potentials are also suitably evaluated
in two schemes, a chiral perturbative approach and an effective meson-quark 
coupling method, with consistent results showing a similar repulsive 
contribution for $K^{+}$ and $K^{0}$. As a consequence we expect rather
reduced effects on the yield ratios. 
We conclude
that particle ratios appear to be robust observables for probing the
nuclear equation of state ($EoS$) at high baryon density and, particularly, 
its isovector sector.
\end{abstract}
\begin{keyword}
Asymmetric nuclear matter, symmetry energy,  
relativistic heavy ion collisions, particle production.\\
PACS numbers: {\bf 25.75.-q}, {\bf 21.65.+f}, 21.30.Fe, 25.75.Dw. 
\end{keyword}
\end{frontmatter}

\date{\today}

\section{Introduction}
The knowledge of the properties of highly compressed and heated hadronic
matter is an important issue for the understanding of astrophysical processes,
such as
the mechanism of supernovae explosions and the physics of neutron
stars \cite{NS1,NS2}. Heavy ion collisions provide the unique opportunity to 
explore highly excited
hadronic matter, i.e. the high density behavior of the nuclear EoS, under
controlled conditions (high baryon energy densities and temperatures)
in the laboratory \cite{dani}. Of particular recent interest is also
the still poorly known density dependence of the
isovector channel of the EoS.

Suggested observables have been the nucleon collective flows
\cite{dani,ritter} and the distributions of produced particles such as pions
and, in particular, particles with strangeness (kaons) 
\cite{fuchs,larionov06}. Because
of the rather high energy threshold ($E_{lab}=1.56$ GeV for Nucleon-Nucleon
collisions), 
 kaon production in HICs at energies in the range $0.8-1.8~AGeV$
is mainly due to secondary processes involving $\Delta$ resonances and pions
($\pi$). On the other hand, secondary processes
require high baryon density. 
This explains why the kaon production around threshold is intimately
connected to the high density stage of the nucleus-nucleus collision.
 Furthermore, the relatively large mean free path
of positive charged ($K^{+}$) and neutral ($K^{0}$) kaons inside the hadronic
environment causes hadronic matter to be transparent for kaons \cite{fuchs06}.
Therefore kaon yields and generally {\it strangeness ratios} have been
proposed as important signals for the investigation of the high density 
behavior of the
nuclear EoS. This idea, as firstly suggested by Aichelin and Ko \cite{ako},
has been recently applied in HIC at intermediate energies in terms of
strangeness ratios, e.g. the ratio of the kaon yields in Au+Au and C+C
collisions \cite{fuchs,oeschler}. In these studies it was found that 
this ratio is
very sensitive to the stiffness of the nuclear EoS. Indeed comparisons 
with KaoS
data \cite{kaos} favored a soft behavior of the high density nuclear EoS, a
statement which is particularly consistent with elliptic flow data of the
FOPI collaboration \cite{fopi}.

The idea of studying particle ratios in HICs around the kinematical threshold 
has been
recently applied in the determination of the isovector channel of the nuclear
EoS, i.e. the high density dependence of the symmetry energy $E_{sym}$. It has
turned out that particle ratios, such as $(\pi^{-}/\pi^{+})$ \cite{bao} or
$(K^{0}/K^{+})$ \cite{stoecker,ferini,ferini2}, are sensitive to the stiffness
of the symmetry energy and, in particular to the strength of the vector 
isovector field. However
in medium effects on the kaon propagation have been
neglected so far. 
Here we will test the robustness of the yield ratio
against the inclusion and the variation of the corresponding kaon potentials. 
At the same time
in Ref. \cite{lari1} the role of the in-medium modifications of NN cross
sections has been studied in terms of baryon and strangeness dynamics. It
was found that the pion and kaon yields are sensitively influenced by the 
reduced
effective NN cross sections for inelastic processes. Here we will see that the 
kaon yield ratio appears robust even with respect to the density dependence
 of the in-medium inelastic NN cross sections, while at variance the pion 
ratio seems to be more sensitive. 

The collision dynamics is rather complex and involves the nuclear mean
field (EoS) and binary $2$-body collisions. In the presence of a nuclear medium
the treatment of binary collisions represents a non-trivial problem.
The NN cross sections for elastic and inelastic processes, which are the
crucial physical parameters here, are experimentally accessible only in
free space and not for $2$-body scattering at finite baryon density. Recent
microscopic studies, based on the $G$-matrix approach, have shown a strong
decrease of the elastic NN cross section \cite{fuchs2,malfliet} in a
hadronic medium. These in-medium effects of the elastic NN cross
section considerably influence the hadronic reaction dynamics \cite{gaitcross}.
Obviously the question arises whether similar in-medium effects of the {\it
inelastic} NN cross sections may affect the reaction dynamics and, in
particular, the production of particles (pions and kaons).

Furthermore,
the strangeness propagation inside the nuclear medium is even more complex and
involves the additional consideration of kaon mean field potentials in the
dynamical description. This is an important issue when comparing with
experimental kaon data \cite{kaos}.  In a Chiral Perturbation approach at
the lowest order (ChPT Potentials), the kaon (antikaon) potential has an 
attractive scalar and a
repulsive (attractive) vector part \cite{kaonpot}. This leads to weakly
repulsive (strongly attractive) potentials for kaons (antikaons) with
corresponding scalar and vector kaon-nucleon coupling constants depending
on the parametrization \cite{kaonpot,kaonpot2} accounted for. 
Similar results can be obtained in an effective meson-coupling model
(OBE Potentials, in the RMF spirit), where the $K$-meson couplings are 
simply related to the 
nucleon-meson ones, in the spirit of ref. \cite{idkpot}. The latter approach 
has the advantage of being fully 
consistent with the covariant transport equations used to simulate the
reaction dynamics \cite{ferini,ferini2}.  
We remind that the high density dependence
of the kaon self energies is still an object of current debate, e.g. see
Refs. \cite{bratko,fuchs06} 
in which the role of the kaon potential has been investigated
in terms of kaon in-plane and out-of-plane flows. Moreover for studies
aimed to the determination of the symmetry energy from strangeness
production one has to consider with particular care the isospin dependence 
of the kaon
mean field potential. 

The main focus of the present work is on a detailed study of the 
{\it robustness} of
the pionic $(\pi^{-}/\pi^{+})$ and, in particular, the strangeness ratio
$(K^{0}/K^{+})$ with respect to the in-medium modifications of the imaginary
part of the nucleon self energy, i.e. the NN cross sections, and to the
in-medium variations of the kaon self energy, i.e. the density
dependence of the kaon potential. 
This analysis, which
goes beyond our previous investigations of \cite{ferini,ferini2}, is also
motivated by new measurements of the FOPI collaboration \cite{lopez} by 
means of
the strangeness ratios.  

The paper is organized as follows: The next Section describes the
theoretical treatment of the reaction dynamics within the Relativistic
Boltzmann-Uheling-Uhlenbeck (RBUU) transport equation. A detailed discussion
on the in-medium modifications of the inelastic NN cross sections
is presented. In Section 3 we discuss the kaon mean field potentials (in both
ChPT and OBE/RMF schemes) and their expected isospin dependence.
Section 4 is devoted to a short introduction to the dynamical calculations.
Results are then shown in Section 5, mostly for central $^{197}Au+^{197}Au$ 
collisions
at $1 AGeV$,
in terms of pion and kaon yields. The initial presentation of the
{\it absolute} yields is relevant for a detailed discussion as well as for 
a comparison with
theoretical results of other groups and with experimental data of the KaoS
and FOPI collaborations. All together this intermediate step is important 
for testing the
reliability of the calculations, since ratios do not do it. Finally
we present the pion and strangeness ratios and discuss their dependence
on the in-medium modifications of the cross NN cross sections 
and of the kaon potentials, including the isospin effects. 
In Section 6 we conclude with a summary and some general comments and 
perspectives.


\section{Theoretical description of the collision dynamics}

In this chapter we briefly discuss the transport equation focusing
on the treatment of two features important for kaon dynamics:
(a) the collision integral by means of the cross sections; 
(b) the kaon mean field potential and its isospin dependence.

\subsection{The RBUU equation}

The theoretical description of HICs is based on the semiclassical
kinetic theory of statistical mechanics, i.e. the Boltzmann
Equation with the Uehling-Uhlenbeck modification of the collision
integral \cite{kada}. The relativistic analog of this equation
is the Relativistic Boltzmann-Uehling-Uhlenbeck (RBUU) equation
\cite{giessen}
\begin{eqnarray}
& & \left[
k^{*\mu} \partial_{\mu}^{x} + \left( k^{*}_{\nu} F^{\mu\nu}
+ M^{*} \partial_{x}^{\mu} M^{*}  \right)
\partial_{\mu}^{k^{*}}
\right] f(x,k^{*}) = \frac{1}{2(2\pi)^9} \nonumber\\
& & \times \int \frac{d^3 k_{2}}{E^{*}_{{\bf k}_{2}}}
             \frac{d^3 k_{3}}{E^{*}_{{\bf k}_{3}}}
             \frac{d^3 k_{4}}{E^{*}_{{\bf k}_{4}}} W(kk_2|k_3 k_4)
 \left[ f_3 f_4 \tilde{f}\tilde{f}_2 -f f_2 \tilde{f}_3\tilde{f}_4
\right] ~,
\label{rbuu}
\end{eqnarray}
where $f(x,k^{*})$ is the single particle distribution function.
In the collision term the short-hand notations $f_i \equiv f(x,k^{*}_i)$
for the particle and $\tilde{f}_i \equiv (1-f(x,k^{*}_i))$
for the hole distributions are used, with 
$E^*_{\bf k} \equiv \sqrt{M^{*2}+{\bf k}^2}$.
The collision integral explicitly exhibits the final state Pauli-blocking while
the in-medium scattering amplitude includes the  Pauli-blocking of intermediate
states.

The dynamics of the drift term, i.e. the lhs of eq.(\ref{rbuu}), is
determined by the mean field. Here the
attractive scalar field $\Sigma_s$ enters via the effective mass
$M^{*}=M-\Sigma_{s}$
and the repulsive vector field $\Sigma_\mu$ via the
kinetic momenta $k^{*}_{\mu}=k_{\mu}-\Sigma_{\mu}$ and via the field tensor
$F^{\mu\nu} = \partial^\mu \Sigma^\nu -\partial^\nu \Sigma^\mu$.
The dynamical description according to Eq.(\ref{rbuu}) involves the strangeness
propagation in the nuclear medium. This topic will be discussed in more
detail at the end of this section.

\subsection{In-medium effects on NN cross sections}

The in-medium cross sections for $2$-body processes (see below) enter in
the collision integral via the transition amplitude
\begin{equation}
W = (2\pi)^4 \delta^4 \left(k + k_{2} -k_{3} - k_{4} \right)
(M^*)^4 |T|^2~~
\label{trans}
\end{equation}
with $T$ the in-medium scattering matrix element. 

\begin{figure}[t]
\unitlength1cm
\begin{picture}(8.,7.3)
\put(3.0,0.3){\makebox{\epsfig{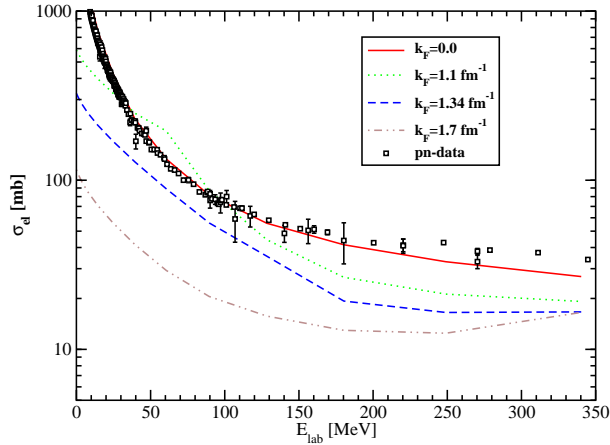}}}
\end{picture}
\caption{Elastic in-medium neutron-proton cross section $\sigma_{el}$ 
at various Fermi momenta
$k_{F}$ as a function of the laboratory energy $E_{lab}$. The free cross
section ($k_{F}=0$) is compared to the experimental total $np$ cross
section \protect\cite{fuchs2}.
}
\label{elastic}
\end{figure}
In the kinetic equation (\ref{rbuu}) both physical input
quantities, the mean field (EoS) and the collision integral (cross sections)
should be derived from the same underlying effective
two-body interaction in the medium,
i.e. the in-medium T-matrix;  $\Sigma \sim \Re T\rho_{B},~~
\sigma\sim \Im T$, respectively.
However, in most practical applications phenomenological mean fields
and cross sections have been used. In such approach the strategy is to
adjust to the known bulk properties
of nuclear matter around the saturation point, and to try to
constrain the models at supra-normal densities
with the help of heavy ion reactions
\cite{dani00,larionov00}. Medium modifications of the NN cross sections
are usually not taken into account. In spite of that for several observables
the comparison to
experimental data appears to work
astonishingly well \cite{dani00,larionov00,gait01,gaitacnm}.
However, in particular kinematical regimes a
sensitivity to the elastic NN cross sections of dynamical observables, 
 such as collective flows and
stopping \cite{gaitcross,gale} or transverse energy transfer
\cite{dancross}, has been observed.

Microscopic Dirac-Brueckner-Hartree-Fock (DBHF) studies for nuclear matter 
above the Fermi energy
regime show a strong density dependence of the elastic
\cite{fuchs2} and inelastic \cite{malfliet,lari2} NN cross sections. In such
studies one starts from the bare NN-interaction in the spirit of
the One-Boson-Exchange (OBE) model by fitting the parameters to
empirical nucleon-nucleus scattering and solves then the 
equations of the nuclear matter many body problem in the
$T$-matrix or ladder approximation. It is not the aim of the
present work to go into further details on this topic. An important feature
of such microscopic calculations is the inclusion of the
Pauli-blocking effect in the {\it intermediate} scattering states
of the $T$-matrix elements and their in-medium modifications, i.e.
the density dependence of the nucleon mass and momenta. Here of
particular interest are the in-medium modifications of the
inelastic NN cross sections since they directly influence the
production mechanism of resonances and thus the creation of pions
and kaons according to the channels listed later (see Sect.3). DBHF 
studies on inelastic NN
cross sections are rare and in limited regions of density and momentum
\cite{malfliet}. For this reason we will first discuss in the following the
in-medium dependence of the elastic NN cross sections, which will be then
used as a starting basis for a detailed analysis of the
density dependence of the inelastic NN cross sections.

The microscopic in-medium dependence of the elastic cross sections
can be seen in Fig. \ref{elastic}, where the energy dependence of the
in-medium neutron-proton $(np)$ cross section at Fermi momenta
$k_F = 0.0, 1.1,1.34,1.7 fm^{-1}$, corresponding to $\rho_{B} \sim
0,0.5,1,2\rho_0$ ($\rho_0=0.16 fm^{-3}$ is the nuclear matter
saturation density) is shown. These results are obtained from
relativistic Dirac-Brueckner calculations \cite{fuchs2}. 
The presence of the medium leads to a substantial
suppression of the cross section which is most pronounced at
low laboratory energy $E_{\rm lab}$ and high densities where
the Pauli-blocking of intermediate states is
most efficient.
At larger $E_{\rm lab}$ asymptotic values of
15-20 mb are reached. 
Also the angular distributions are affected by the presence of the
medium. E.g. the initially strongly forward-backward
peaked  $np$ cross sections become much more isotropic at finite densities,
 mainly due to the Pauli suppression of intermediate
soft modes ($\pi$-exchange) \cite{fuchs2}. As a consequence a larger
transverse energy transfer can be expected.

The case of the inelastic NN cross sections is similar, but more complicated. 
The presence of the medium influences not only the matrix elements, but 
also the threshold energy $E_{\rm tr}$, which is an important quantity at 
beam energies below or near the threshold of particle production.  
In free space it is calculated from the invariant 
quantity $s=(p_{1}^{\mu}+p_{2}^{\mu})(p_{1\mu}+p_{2\mu})$ with 
$p_{i}^{\alpha},~(i=1,2)$ the $4$-momenta of the two particles in the 
ingoing collision channel, e.g. $NN \longrightarrow N\Delta$. This 
quantity is conserved in binary collisions in free space, 
from which one determines the modulus of the momenta of the particles 
in the outgoing channel. The threshold condition reads 
$E_{\rm tr} \equiv \sqrt{s} \ge M_{1}+M_{2}$. Cross sections in free 
space are usually parametrized in terms of $\sqrt{s}$ or the 
corresponding momentum in the laboratory system $p_{lab}$ within the 
One-Boson-Exchange (OBE) model, see e.g. \cite{huber} for details. 

At finite density, however, particles carry kinetic momenta and 
effective masses and obey a dispersion relation 
$p_{\mu}^{*}p^{*\mu}=m^{*2}$ modified with respect to the free case. 
These in-medium effects shift the threshold energy in the free space according 
 to $s^{*}=(p_{1}^{*\mu}+p_{2}^{*\mu})(p_{1\mu}^{*}+p_{2\mu}^{*})$ and the 
threshold condition for inelastic processes inside the medium reads now 
$E^{*}_{\rm tr} \equiv \sqrt{s^{*}} \ge m_{1}^{*}+m_{2}^{*}$. The 
requirement of energy-momentum conservation can be carried out in terms 
of the quantity $s^{*}$ $or$ $s$, $only$ as long as the in-medium mean 
fields or the 
corresponding self energies do not change between ingoing and outgoing 
channels. 

\begin{figure}[t]
\unitlength1cm
\begin{picture}(8.,7.3)
\put(2.0,0.3){\makebox{\epsfig{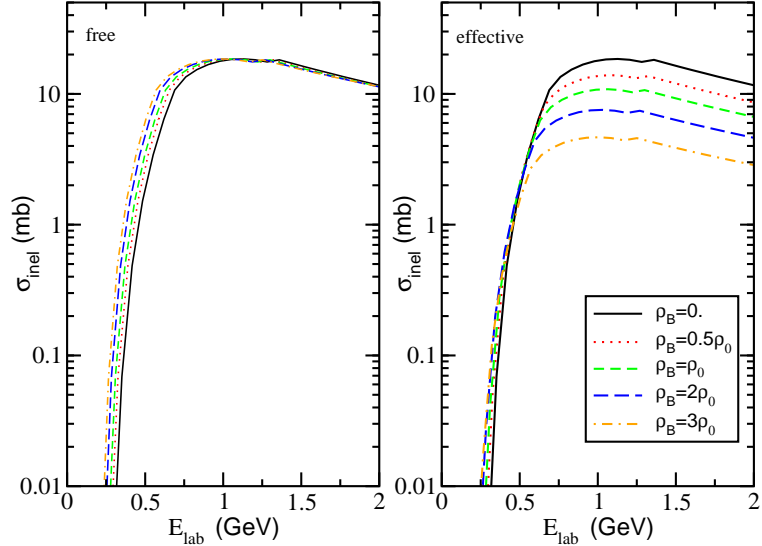}}}
\end{picture}
\caption{Inelastic $NN\longrightarrow N\Delta$ cross section 
$\sigma_{inel}$ at various baryon densities $\rho_{B}$ (in units of the 
saturation density $\rho_{0}=0.16~fm^{-3}$) as a function of 
the laboratory energy $E_{\rm lab}$ using the free parametrizations
 (left) and the
in-medium modified ones from 
DBHF \protect\cite{malfliet} (right).
}
\label{inelastic}
\end{figure}

The application of free parametrizations of cross sections for inelastic 
processes in dynamical situations of HICs at finite 
density leads thus to an inconsistency, since the threshold condition is 
performed in terms of effective quantities, but the matrix elements are 
carried out in free space, e.g. by fitting their parameters to free 
empirical NN scattering. This effect can be seen in 
Fig. \ref{inelastic} (left panel) where the free inelastic 
$NN \longrightarrow N\Delta$ cross section $\sigma_{inel}$ as a function of 
the laboratory energy $E_{\rm lab}$ is displayed, at various baryon 
densities $\rho_{B}$. The threshold energy in the free space is 
$E_{\rm tr} = \sqrt{s}=2.014$ GeV (for $M=0.939$ GeV and $M_{min}=1.076$ 
for the 
nucleon and the lower limit mass of the $\Delta$ resonance). The corresponding 
threshold value of the laboratory energy 
$E_{\rm lab}=(E_{\rm tr}^{2}-4M^{2})/2M$ is $0.32$ GeV. However, at finite 
density 
the threshold is shifted towards lower energies, i.e. the free cross section 
increases, due to the reduction of the free masses of the outgoing particles 
in the threshold condition $E^{*}_{\rm tr} \ge m^{*}_{1}+m^{*}_{2}$. Obviously 
at higher energies far from threshold the free cross section does not depend 
on the density. 

A more consistent approach is the determination of the inelastic 
cross section under the consideration of in-medium effects, i.e. 
the Pauli-blocking of intermediate scattering states and in-medium modified 
spinors in the determination of the matrix elements within the OBE model. 
A simultaneous treatment of the transport equation and the structure
equations of DBHF for actual anisotropic momentum configurations is not
possible, due 
to its high complexity. For this reason we have applied the same 
method as for the case of elastic binary processes, i.e. in-medium 
parametrizations of the {\it inelastic} cross sections of the type 
$NN \longrightarrow N\Delta$ within the same underlying DBHF approach as 
already used for
the elastic processes. 
Haar and Malfliet \cite{malfliet} investigated this topic for infinite nuclear
matter with the result of a strong in-medium modification of the inelastic
cross sections due to the reasons given above. However, these studies were 
performed at various densities but only in a limited region of momenta. 
For a practical application in 
HICs we have thus extended these 
DBHF calculations using an extrapolation technique.
We have imposed an exponential decay law of the form $ae^{-bp_{lab}}$
on the values of the in-medium cross sections of the channel
$NN \rightarrow N\Delta$ given in ref. \cite{malfliet}. The parameter $a$
normalizes to the last value of the extrapolated cross section and $b$ is 
defined by fitting the slope of the free cross section, since it does not 
change with density. For the density dependence we have enforced a correction 
of the
form $f(\rho_{B})=1+a_0(\rho_{B}/\rho_0)+a_1(\rho_{B}/\rho_0)^2+a_2(\rho_{B}/\rho_0)^3$,
 where $a_0=-0.601,a_1=0223,a_2=-0.0035$, with $\rho_0$ saturation density,
are extracted from the results of ref.\cite{malfliet}.
The same modification is imposed on the cross sections of all the inelastic 
channels,
in a form of the type 
$\sigma_{eff}=\sigma_{free}(E_{\rm lab}) f(\rho_{B})$, with $\sigma_{free}$ taken 
from the standard free parametrizations of Ref. \cite{huber}. Such a procedure 
is well appropriate at low energies but at higher momenta can be less 
accurate. This,  however, should not be a problem at the reaction energies 
below the kaon production threshold considered in this work. 

Fig. \ref{inelastic} (right panel) shows the energy dependence of the 
inelastic NN
cross section at various densities as obtained from DBHF
calculations \cite{fuchs2} for symmetric nuclear matter. 
 As in the case of elastic processes (see Fig.
\ref{elastic}), the inelastic one drops with increasing baryon
density $\rho_{B}$ mainly due to the Pauli blocking of intermediate
scattering states and the in-medium modification of the effective
Dirac mass \cite{fuchs2}. 
There are also phenomenological 
studies \cite{lari1,lari2} which give similar medium effects on
the inelastic cross sections, within the limitation to isospin symmetric 
nuclear matter. 
More suitable results would come from a
 DBHF approach to isospin asymmetric nuclear matter. Only
recently such studies have been started \cite{fuchs3}, however,
limiting to low momenta regions, below the threshold energy of
inelastic channels.

\section{Kaon Potentials}

Before starting with the presentation of the results, it is
important to analyse the in-medium kaon potential, since
it could be relevant when theoretical results will be compared with
experiments. In fact it has been widely discussed whether
the kaon potential plays a crucial role in describing kaon
production and their dynamics \cite{bratko,fuchs06,oeschler}. Kaplan and Nelson
\cite{kaonpot} found that the explicit chiral symmetry breaking is not
so small for $K$ mesons and this leads to significant corrections to
the free kaon mass at finite baryon density. There are different
models for the description of kaon properties in the nuclear medium.
Here we will briefly discuss two main approaches, one based on Chiral 
Perturbation Theory ($ChPT$) and a second on effective meson couplings 
 ($OBE/RMF$), more 
consistent with
the general frame of our covariant reaction dynamics. The results are in 
good agreement and this is not surprising on the basis of a simple physics
 argument.
It is well established \cite{fuchs06} that 
kaons
($K^{0,+}$) feel a weak repulsive potential in nuclear matter, of
the order of $20-30~MeV$ at normal density. This can be described as the net 
result of the cancellation of
an attractive  scalar 
and a repulsive vector interaction terms.
Such a mechanism can be reproduced in the ChPT approach through  the 
competition between
an attractive  scalar Kaplan-Nelson term \cite{kaonpot}
and a repulsive vector Weinberg-Tomozawa \cite{kaonpot3}
term. The same effect can be obtained in an effective meson field scheme
just via a coupling to the attractive $\sigma$-scalar and to the repulsive
$\omega$-vector fields.

In this paper antikaons
$K^{-}$ and their strong attractive potential will be not discussed, 
since 
for the higher threshold they have been not considered in the energy range
of interest here.

Finally, for studies
aimed to the determination of the symmetry energy from strangeness
production one has to treat with particular care the isospin dependence 
of the kaon
mean field potential.
\subsection{Chiral Perturbative Results}
\begin{figure}[t]
\unitlength1cm
\begin{picture}(8.,7.)
\put(2.0,0.0){\makebox{\epsfig{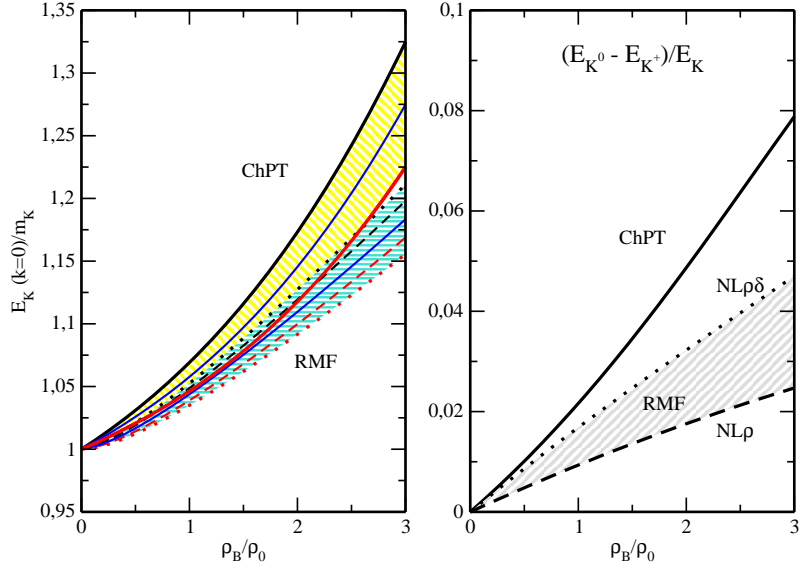}}}
\end{picture}
\caption{Density dependence ($\rho_0$ is the saturation density)
of in medium kaon energy (left panel)
in unit of the free kaon mass ($m_K=0.494\, GeV$).
Upper curves refer to $ChPT$ model calculations: the central line corresponds 
to symmetric matter, the other two give the isospin effect (up $K^0$, down
$K^+$). Bottom curves are obtained in the $OBE/RMF$ approach, the solid one 
is for 
symmetric matter. The isospin splitting is given by the dashed ($NL\rho$)
and dotted ($NL\rho\delta$) lines, again up $K^0$, down $K^+$.
Right panel: relative weight of the isospin splitting, see text. 
 All the curves are obtained considering an asymmetry
parameter $\alpha=0.2$.} \label{kmass}
\end{figure}

Starting from an effective chiral Lagrangian for the $K$
mesons one obtains a density and isospin dependence for the
effective kaon ($K^{0,+}$) masses \cite{fuchs06}.
In isospin asymmetric matter we finally get
\begin{equation}
m_K^* =\sqrt{m_K^2 -\frac{\Sigma_{KN}}{f_\pi^2}\rho_s \mp
\frac{C}{f_\pi^2}\rho_{s3}+ V_\mu V^\mu} ~~~~~~(upper~sign,~K^+),
\label{kaoneffiso}
\end{equation}
where $\rho_s,~\rho_{s3}$ are total and isospin scalar densities,
with $m_K=494 MeV$ the free kaon mass, $f_\pi=93 MeV$ the pion decay constant,
and $\Sigma_{KN}$ the kaon-nucleon sigma term (attractive scalar), here
chosen as 450 MeV.
The vector potential is given by:
\begin{equation}
V_\mu = \frac{3}{8{f^*}_\pi^2}j_\mu \pm
\frac{1}{8{f^*}_\pi^2}j_{\mu3}~~~~~~~~~~(upper~sign,~K^+), 
\label{kapotiso}
\end{equation}
with $j_\mu,~j_{\mu3}$ baryon and isospin currents.
The $f^*_\pi$ is an in-medium reduced pion decay constant. It is expected to
scale with density in a way similar to the chiral condensate \cite{BroRho}. 
This leads to a reduction around normal density 
 ${f^*}_\pi^2 \simeq 0.6 f_\pi^2$.
Such a reduction is compensated in one-loop ChPT by other contributions
 in the scalar 
attractive term so we will use $f^*_\pi$ only for the vector
potential, with an enhanced repulsive effect \cite{fuchs06}.
The constant $C$ has been fixed from the Gell-Mann-Okubo mass formula 
(i.e. in free space) to a value of
$33.5 \, MeV$ \cite{idkpot}.
In Eqs. (\ref{kaoneffiso}-\ref{kapotiso}) upper signs hold for
$K^{+}$ and lower signs for $K^{0}$. 
As can be seen, the vector term,
which dominates over the scalar one at high density, is more
repulsive for $K^0$ than for $K^+$. This leads to a higher (lower)
$K^0$ ($K^+$) kaon in-medium energy given by the dispersion relation
\begin{equation}
E_K({\bf k}) = k_0 = \sqrt{{\bf k}^2 + {m^*}_K^2} + V_0
\label{energy}
\end{equation} 
The density dependence, evaluated in the chiral approach, of
the quantity $E_K({\bf k})_{{\bf k}=0} = m^*_K + V_0$ for $K^{0,+}$, that 
directly influences the in-medium production thresholds is shown by the 
upper curves in Fig. \ref{kmass} (left panel). In particular, it can be 
noted that
that $K^0$ and $K^+$ in medium-energy differs by $\approx 5\%$ at
$\rho_{B}=2\rho_0$ (with $E_{K^0}>E_{K^+}$), at a 
fixed isospin asymmetry around $0.2$. Therefore, the 
inclusion of isovector terms
favors $K^+$ over $K^0$ production, with a consequent reduction of
the $K^0/K^+$ strangeness ratio.

\subsection{Relativistic Mean Field Results}
Kaon potentials can be also derived within an effective meson field 
$OBE$ approach,
 fully consistent with the $RMF$ transport scheme used to simulate the 
reaction dynamics, see Eq.(\ref{rbuu}). We will use a simple constituent 
quark-counting
prescription to relate the kaon-meson couplings to the nucleon-meson 
couplings, i.e. just a factor 3 reduction. Following the chiral argument 
discussed before,
 only for the scalar vector case we have further increased the kaon coupling
 to $g_{\omega K} \simeq 1.4/3 g_{\omega N}$. This will ensure the required
repulsion around normal densities for $K^+$s.
Consistently the isospin dependence will be directly derived from
the coupling between the kaon fields and the $\rho$ and $\delta$
isovector mesons \cite{idkpot}.

The in-medium energy carried by kaons will have the same form as in
Eq.(\ref{energy}) but with effective masses and vector potentials given by
\begin{eqnarray}
m_K^* =\sqrt{m_K^2 - m_K (g_{\sigma K} \sigma \pm 
\frac{g_{\delta K}g_{\delta N}}{m_\delta^2} \rho_{s3})} \nonumber \\
= \sqrt{m_K^2 - m_K (g_{\sigma K} \sigma \pm \frac{1}{3} f_\delta \rho_{s3})} 
\label{massrmf}
\end{eqnarray}
\begin{eqnarray}
V_0 = \frac{g_{\omega K}g_{\omega N}}{m_\omega^2} \rho_{B} \pm 
 \frac{g_{\rho K}g_{\rho N}}{m_\rho^2} \rho_{B3} \nonumber \\
 = \frac{1}{3} ({{f_\omega}^*}\rho_B \pm f_\rho \rho_{B3})
\label{vectormf}
\end{eqnarray}
where upper signs are for $K^+$s. The 
$f_i \equiv g_{iN}^2/m_i^2,~i=\sigma,\omega,\rho,\delta$ are the nucleon-meson 
coupling constants used in our $RMF$ Lagrangians and 
$f_\omega^*=1.4 f_\omega$ due to the enhanced kaon-scalar/vector coupling.
$\sigma$ represents the solution of the non linear equation for the 
scalar/isoscalar field which gives the reduction of the nucleon 
mass in symmetric matter, therefore we can directly evaluate the
 kaon-$\sigma$ coupling using
$$
g_{\sigma K} \sigma = \frac{1}{3}(M - M^*)
$$                    
where $M^*$ is the nucleon effective mass at the fixed baryon density.

In this $RMF$ approach we can derive an almost analytical expression for
 the isospin
 effects on the kaon in-medium energy Eq.(\ref{energy}) at ${\bf k} = 0$. 
Using the
approximate form $\rho_s \simeq M^*/E_F^* \rho_B$ for the scalar density,
we get a relative weight of the isospin splitting of the kaon potentials
$\Delta E_K({\bf k})_{{\bf k}=0} \equiv E_{K^0}({\bf k})_{{\bf k}=0} - 
E_{K^+}({\bf k})_{{\bf k}=0}$
given by
\begin{equation}
\frac{\Delta E_K}{E_K}= \frac{2\alpha(f_\rho - \frac{M^*}{2E^*_F}f_\delta)}
{f^*_\omega + \frac{3}{\rho_B} (m_K - \frac{1}{6}(M - M^*))}
\label{isosplit}
\end{equation}
with $\alpha \equiv \rho_{B3}/\rho_B$ the asymmetry parameter.

We can now easily estimate the isospin splitting of $K^0$ vs. $K^+$ for the 
two isovector mean field
Lagrangians used here, $NL\rho$ and $NL\rho\delta$. The effect will be clearly
larger when the $\delta$ coupling is included since we have to increase
the $\rho$-coupling $f_\rho$, see \cite{ferini,ferini2}, but still the expected
weight is relatively small, going from about $1.5 \%$ ($NL\rho$) to about
$3.0 \%$ ($NL\rho\delta$) at
$\rho_B=2\rho_0$, for a fixed isospin asymmetry around $0.2$.
 The complete results are also shown in Fig. \ref{kmass} (right panel).
The agreement with the ChPT estimations is rather good, but in the $RMF$ scheme
we see an overall reduced repulsion and a smaller isospin splitting. Both 
effects are of interest for our discussion, the first affecting the $K^{0,+}$
 absolute yields, the second important for the $K^0/K^+$ yield ratios.

\section{Numerical realization and notations}

The Vlasov term of the RBUU equation (\ref{rbuu}) is treated within the 
Relativistic Landau-Vlasov method, in which the phase space distribution 
function $f(x,p^{*})$ is represented by covariant Gaussians in coordinate 
and momentum space \cite{RLV}. For the nuclear mean field or the 
corresponding EoS in symmetric matter the $NL2$ parametrization 
\cite{giessen} of the 
non-linear Walecka model \cite{walecka} is adopted with a compression
modulus of $200$ MeV and a Dirac effective mass of $m^{*}=0.82~M$
($M$ is the bare nucleon mass) at saturation. The momentum dependence 
enters via
the relativistic treatment in terms of the vector component of
the baryon self energy. 
The isovector components in the mean fields are introduced in the
$NL\rho,NL\rho\delta$ Lagrangians as in the recent Refs. 
\cite{ferini,ferini2}. In Table \ref{table1} we report all the coupling 
constants and the coefficients of the non-linear $\sigma$-terms.
\nopagebreak
\begin{table}[t]
\begin{center}
\begin{tabular}{|l|c|c|c|c|c|c|c|c|c|}
\hline\hline 
       & $f_{\sigma}$ ($fm^2$)   & $f_{\omega}$ ($fm^2$) & $f_{\rho}$ ($fm^2$)
     & $f_{\delta}$ ($fm^2$)     & A ($fm^{-1}$) & B &  \\ 
\hline\hline
   $NL\rho$          &     9.3        &     3.6     & 1.22 &    0.0     &
  0.015     &    -0.004 &  \\ 
\hline
   $NL\rho\delta$ &     9.3        &     3.6    & 3.4 &    2.4     &
  0.015     &    -0.004 &  \\ 
\hline\hline
\end{tabular}
\end{center}
\vskip 0.5cm
\caption{\label{table1} 
Coupling parameters in terms of $f_{i} \equiv 
(\frac{g_{i}}{m_{i}})^{2}$ 
for $i=\sigma,~\omega,~\rho,~\delta$, 
$A \equiv \frac{a}{g_{\sigma}^{3}}$ and 
$B \equiv \frac{b}{g_{\sigma}^{4}}$ for the non-linear $NL$ models 
\cite{ferini} 
using the $\rho$ ($NL\rho$) and 
both, the $\rho$ and $\delta$ mesons ($NL\rho\delta$) for the 
description of the isovector mean field.}
\vskip 0.5cm
\end{table}

The collision integral 
is treated within the standard parallel ensemble algorithm imposing 
energy-momentum conservation. For the elastic NN cross sections the 
DBHF calculations of Ref. \cite{fuchs2} have been used throughout this work. 
At intermediate relativistic energies up to the threshold of kaon 
($K^{0,+}$) production, i.e. $E_{lab}=1.56$ GeV, the major inelastic 
channels are ($B,Y,K$ stand for a baryon (nucleons $N$ or 
a $\Delta$-resonance), hyperon and kaon, respectively)
\begin{itemize}
\item $NN \longleftrightarrow N\Delta$ ($\Delta$-production and absorption)
\item $\Delta \longleftrightarrow \pi N$ ($\pi$-production and absorption)
\item $BB \longrightarrow BYK$, $B\pi \longrightarrow YK$ 
($K$-production from $BB$ and $B\pi$-channels)
\end{itemize}
The produced resonances propagate in the same mean field as the nucleons, and 
their decay is characterized by the energy dependent lifetime $\Gamma$ which 
is taken from Ref. \cite{huber}. The produced pions 
propagate under the influence 
of the Coulomb interaction with the charged hadrons. Kaon production is 
treated hereby perturbatively due to the low  cross 
sections, taken from Refs. \cite{tsushima}. Kaons undergo 
elastic scattering and 
their phase space trajectories are determined by relativistic equations of 
motion, if the kaon potential is accounted for. 

In the next section the results of transport calculations in terms of 
pion and kaon yields and their rapidity distributions will be presented. 
The following cases for the inelastic NN cross sections $\sigma_{inel}$ and 
the kaon potential $\Sigma_{K}$ (scalar and vector) will be particularly 
discussed: 
\begin{itemize}
\item free $\sigma_{inel}$, without $\Sigma_{K}$
(w/o K-pot $\sigma_{free}$)
\item free $\sigma_{inel}$, with $\Sigma_{K}$
(w K-pot $\sigma_{free}$)
\item free $\sigma_{inel}$, with isospin dependent $\Sigma_{K}$
(w ID K-pot $\sigma_{free}$)
\item effective $\sigma_{inel}$, without $\Sigma_{K}$
(w/o K-pot $\sigma_{eff}$)
\item effective $\sigma_{inel}$, with $\Sigma_{K}$
(w K-pot $\sigma_{eff}$)
\item effective $\sigma_{inel}$, with isospin dependent $\Sigma_{K}$
(w ID K-pot $\sigma_{eff}$)
\end{itemize}
For pions only the different cases of $\sigma_{inel}$ will 
be labelled, since they do not experience any  potential, apart coulomb. 
One should 
note that in all calculations only inelastic processes including 
the lowest mass resonance $\Delta(1232 MeV)$ have been considered, without 
accounting for the $N^{*}(1440)$ resonance. This will have not appreciable
consequences for pions yields, but it slightly reduces the kaon 
multiplicities. 

\section{Results}

As mentioned in the introduction, the main topic of the present work is 
to study the sensitivity of particle ratios to physical parameters such as 
in-medium effects of cross sections and the isospin dependence of the 
kaon potential. This is an important issue to clarify since there is some 
evidence suggesting the
yield ratios as good observables in determining the high density 
behavior of the symmetry energy. In a near future these data will be 
experimentally accessible with the help of reactions 
with radioactive ion beams. However, a comparison of absolute values with 
experimental data, although it is not the aim of this work, is essential and 
it has to be included in order to show the consistency of our approach. 
Thus we will start the presentation of the results first in terms of absolute 
yields, and comparison with data, before passing to the main section on 
the particle ratios. Most calculations refer to central $^{197}Au+^{197}Au$
collisions at $1~AGeV$.

\subsection{Effects of in-medium inelastic NN cross sections on 
particle yields}
\subsubsection{Resonance and Pion Production}
\begin{figure}[t]
\unitlength1cm
\begin{picture}(8.,7.)
\put(3.0,0.3){\makebox{\epsfig{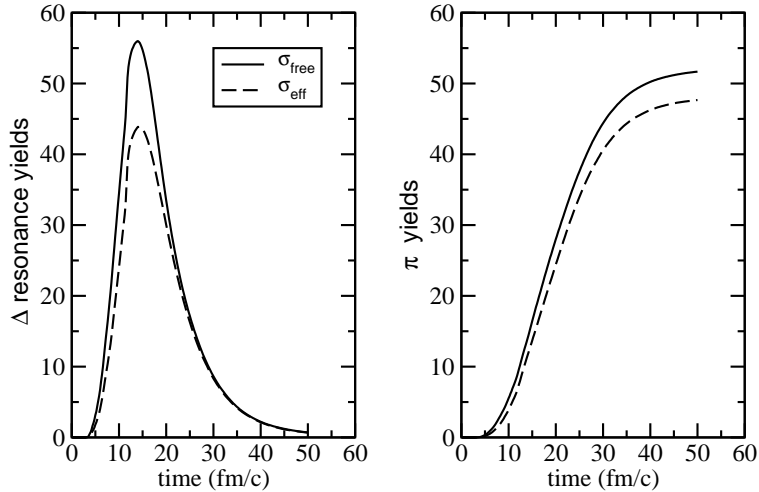}}}
\end{picture}
\caption{Time evolution of the $\Delta$-resonances (left panel) and
total pion yield (right panel) for a central ($b=0~fm$) Au+Au reaction at
1 AGeV incident energy. Calculations with free (solid lines) and
effective (DBHF, dashed lines) $\sigma_{inel}$ are shown. 
}
\label{deltapions}
\end{figure}
Here we study the role of the density dependence of the effective 
inelastic NN cross sections on particle yields (pions and kaons). 
We start with the temporal evolution of the $\Delta$ resonances and 
the produced pions, as shown in Fig. \ref{deltapions}. 
The maximum of the multiplicity of produced $\Delta$-resonances
occurs around $15$ fm/c which corresponds to the time of
maximum compression. Due to their finite lifetimes these resonances decay into
pions (and nucleons) as $\Delta \longrightarrow \pi N$. Some of these
pions are re-absorbed in the inverse process,
i.e. $\pi N \longrightarrow \Delta$ but $chemical~equilibrium$ is never
reached, as pointed out in \cite{ferini2}. This mechanism continues until all
resonances have decayed leading to a saturation of the pion yield
for times $t \geq 50$ fm/c (the so-called freeze-out time). 
The resonance production 
takes place during the high density phase, where the in-medium effects of 
the effective cross sections are expected to dominate. In fact, the 
transport results with the in-medium modified $\sigma_{inel}$ 
reduce the multiplicity of inelastic processes, and thus the yields of 
$\Delta$ resonances and pions. 
However, the in-medium effect is not so pronounced here with respect to 
similar phenomenological studies of Ref. \cite{lari1,lari2}, which should come 
from the moderate density dependence of the effective cross sections, see 
also again Fig. \ref{inelastic}.

\begin{figure}[t]
\unitlength1cm
\begin{picture}(8.,10.)
\put(3.0,0.3){\makebox{\epsfig{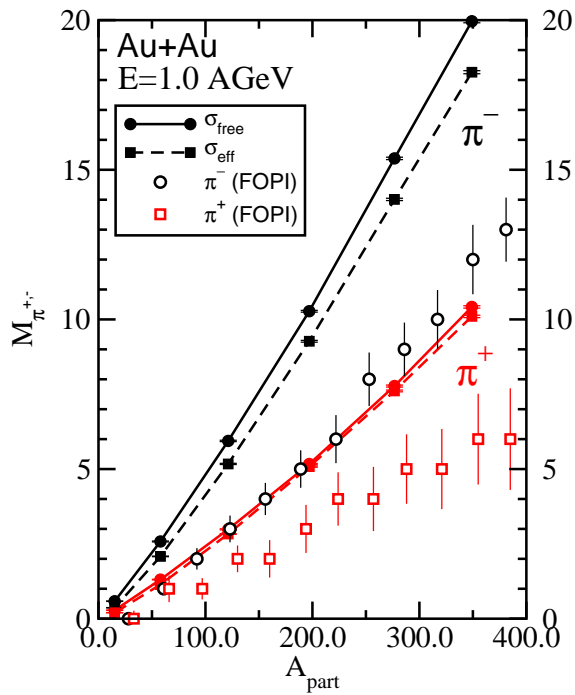}}}
\end{picture}
\caption{ Centrality dependence (in terms of $A_{part}$) of the 
negative ($\pi^{-}$) and positive ($\pi^{+}$) charged pions for 
Au+Au collisions at 1 AGeV incident energy. Calculations with 
free (solid lines, filled circles) and effective (dashed lines, 
filled squares) cross sections are shown as indicated. Experimental 
data, taken from FOPI collaboration \cite{pelte}, are also displayed for 
comparison. 
}
\label{picentr}
\end{figure}

Fig. \ref{picentr} shows the centrality dependence of the charged pion yields 
for Au+Au collisions at 1.0 AGeV incident energy. The degree of centrality 
is characterized by the observable $A_{part}$, which gives 
the number of participant nucleons and can be calculated within a 
geometrical picture using smooth density profiles for the nucleus 
\cite{pelte}. Obviously $A_{part}$ increases with decreasing impact 
parameter $b$ and its value approaches the total mass number of the two 
colliding nuclei in the limiting case of $b=0$ fm. As can be seen in 
Fig. \ref{picentr}, the charged pion yields are enhanced with increasing 
$A_{part}$, particularly in a non-linear $A_{part}$-dependence. As pointed 
out in  \cite{pelte}, the charged pion multiplicities show
a similar non-linear increase  
also in the data. However, by directly comparing the 
theoretical charged pion yields with the experiments 
\cite{pelte} 
we observe that our calculations overpredict the data, even when the  
in-medium reductions in $\sigma_{inel}$ are accounted for.

This discrepancy is a general feature of the transport models and may lie on 
the role of the 
rescattering processes that take place in the spectator region, where
nuclear surface effects can play a crucial role. In order to check this point
we have performed a selection on pions produced at central rapidity, where 
data are also available \cite{pelte}.
 
\begin{figure}[t]
\unitlength1cm
\begin{picture}(8.,7.)
\put(2.0,0.){\makebox{\epsfig{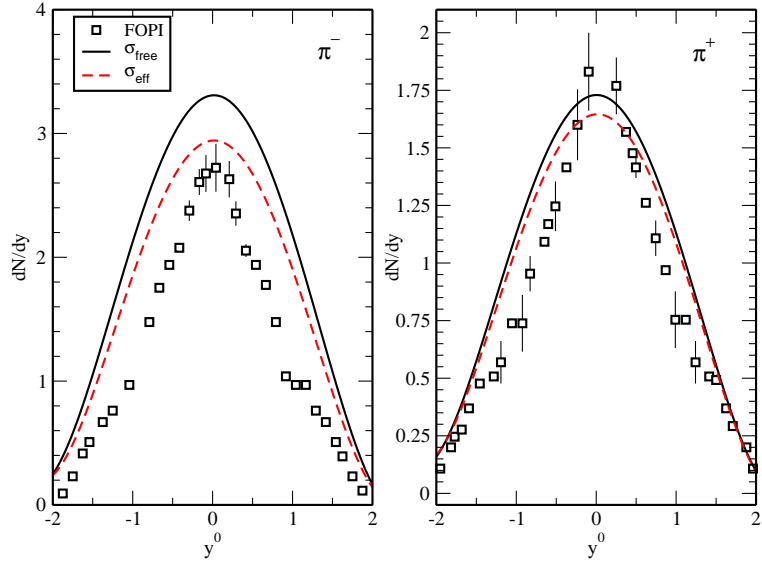}}}
\end{picture}
\caption{Inclusive (all centralities) pion rapidity distributions
for a Au+Au reaction 
at $E_{\rm beam}=1$ AGeV incident energy. Comparison with the experimental 
values 
given by FOPI collaboration \cite{pelte}; as in the data we have used a
transverse momentum cut to $p_t > 0.1 GeV/c$. 
}
\label{pirapincl}
\end{figure}

In Fig.~\ref{pirapincl} we present the inclusive (all centralities) pion 
rapidity distributions vs. the FOPI data for charged pions. We see that the
agreement is rather good at mid-rapidity while we see a definite overcounting
in the spectator sources.  

\begin{figure}[t]
\unitlength1cm
\begin{picture}(8.,7.)
\put(2.0,0.){\makebox{\epsfig{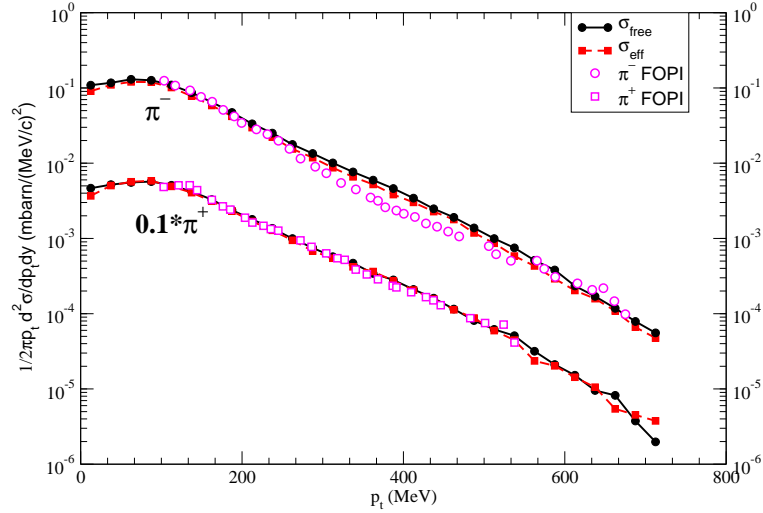}}}
\end{picture}
\caption{Inclusive transverse spectrum at midrapidity of $\pi^{-}$, $\pi^{+}$ 
for a Au+Au reaction 
at $E_{\rm beam}=1$ AGeV incident energy. Comparison with the experimental 
values 
given by FOPI collaboration \cite{pelte}. The cross sections are normalized to
 a rapidity interval $dy=1$.
}
\label{pispectr}
\end{figure}
Such a good evaluation of the pion production ad mid-rapidity is confirmed
by the results shown in
Fig.~\ref{pispectr}, where we present the inclusive (all centralities) pion 
transverse spectrum 
at midrapidity ($-0.2<y^{0}<0.2$). We first note that this is also not much 
 affected by the inclusion of
the in-medium inelastic cross sections. Moreover we see again that
our results are in good agreement with the 
experimental
values from the FOPI collaboration \cite{pelte}, in the same rapidity 
selection.
The overestimation of the pion yields shown in
Fig.~\ref{picentr}  probably results from 
other rapidity regions where the role of the spectator sources is more evident.
We have also to say that we are not imposing any experimental filter to our 
results. The point is rather delicate since the main discrepancies appear 
in high rapidity regions.
In any case such a fine agreement at mid-rapidity is very important for the 
reliability of our results on kaon production, mostly produced in that 
rapidity range via secondary $\pi N,\Delta N$ channels, see \cite{ferini2}. 
 
\begin{figure}[t]
\unitlength1cm
\begin{picture}(8.,7.)
\put(3.0,0.3){\makebox{\epsfig{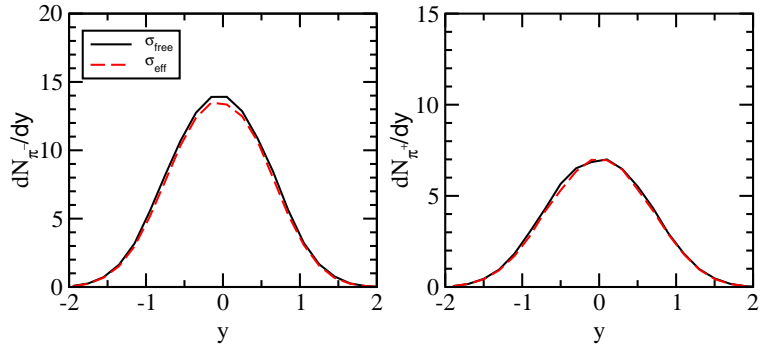}}}
\end{picture}
\caption{Rapidity distributions of negative and positive 
charged pions (left and right panels, respectively) for 
a central ($b=0$ fm) Au+Au reaction 
at $E_{\rm beam}=1$ AGeV incident energy.
}
\label{pirap}
\end{figure}
The pion reaction dynamics is furthermore not sensitively affected
by the in-medium inelastic cross sections. We restrict here the analysis
to $central$ Au+Au collisions at 1 AGeV. 
In Fig. \ref{pirap} we show cross section effects on the 
rapidity distributions 
(normalized to the projectile rapidity in the $cm$ system) for 
$\pi^{\pm}$, an observable which characterizes the degree of stopping or the 
transparency of the colliding system. This is due to the fact that the 
global dynamics is mainly governed by the total NN cross sections, in which 
its elastic contribution is the same for all the cases. In previous 
studies \cite{gaitcross,gale} the in-medium effects of 
the {\it elastic} NN cross 
sections gave important contributions to the degree of transparency or 
stopping. It was found that a reduction of the effective NN cross section 
particularly at high densities is essential in describing the experimental 
data \cite{gaitcross}, as confirmed by various other analyses \cite{gale}. 
The density effects on the inelastic NN cross section influence only those 
nucleons associated with resonance production, and therefore they do not 
affect the global baryon dynamics significantly. 

\subsubsection{Kaon Production}

\begin{figure}[t]
\unitlength1cm
\begin{picture}(8.,7.)
\put(3.0,0.0){\makebox{\epsfig{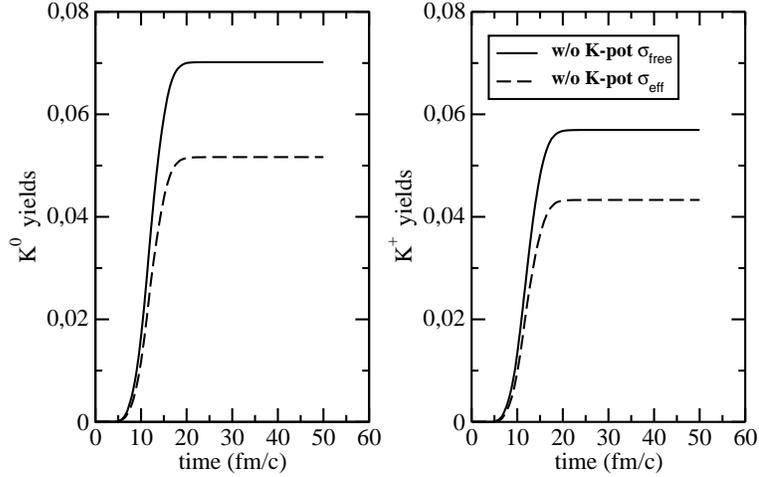}}}
\end{picture}
\caption{Time evolution of the $K^{0}$ (left panel) and $K^{+}$ (right panel)
multiplicities, for the same reaction and models as in 
Fig.~\protect\ref{deltapions}, with free and in-medium inelastic cross
sections, without 
the inclusion of the kaon potentials.
}
\label{kaons}
\end{figure}

The situation is different for kaon production, see Fig. \ref{kaons}. The 
influence of the in-medium dependence of $\sigma_{inel}$ is important, 
and reduces the kaon abundancies by a factor of $\approx 30\%$. This is due to
the fact that the leading channels for kaon production are
$N\Delta \longrightarrow BYK$ and $N \pi \longrightarrow \Lambda K$. 
Thus kaon production is essentially a twostep process and the
medium-modified inelastic cross sections enter twice, leading to an
increased sensitivity.

\begin{figure}[t]
\unitlength1cm
\begin{picture}(8.,7.)
\put(2.0,0.0){\makebox{\epsfig{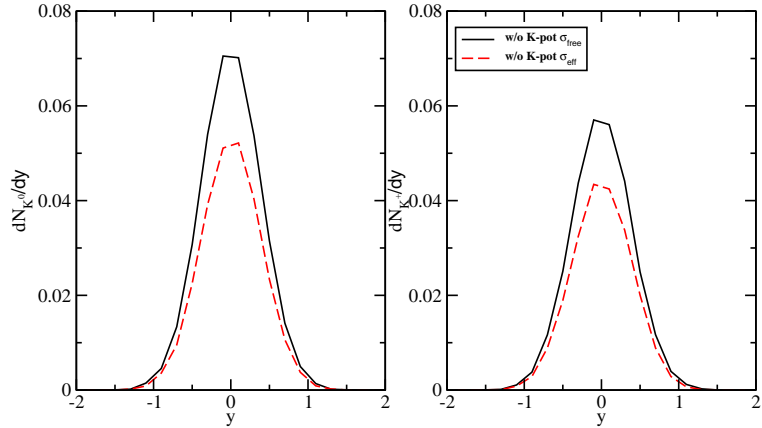}}}
\end{picture}
\caption{Same as in Fig. \protect\ref{kaons}, but for the normalized 
rapidity distributions.
}
\label{krap}
\end{figure}

Fig.~\ref{krap} shows the rapidity distributions of kaons, 
where the in-medium effect is more visible with respect to the 
corresponding pion rapidity distributions (see Fig.~\ref{pirap}). 
These results seem to show that kaon production could be used to 
determine the in-medium dependence of the NN cross section for inelastic 
processes. Similar phenomenological studies based on the BUU approach 
\cite{lari1,lari2} strongly support in-medium modifications of the free cross 
sections. It is of great interest to perform an extensive comparison with 
experimental data on kaon production, in order to have a more clear image of
the effect of the in-medium cross sections on their production.
The point is that kaon absolute yields are also largely affected
by the kaon potentials, see the following, as expected from the general 
discussion of the previous section. However since kaons are mainly produced
in more uniform high density regions the effects of the medium on cross 
sections 
tend to disappear in the yield ratios. In the next section we will show that 
the same holds true for the $K^0, K^+$ potentials. Our conclusion is that the
kaon yield ratios might finally be a rather $robust$ observable to
probe the nuclear $EoS$ at high baryon densities.

\subsection{The role of the kaon potential}

As discussed in the previous sections, the important quantity which influences 
the kaon production threshold is the in-medium energy at zero momentum 
\cite{fuchs06}. This quantity rises with increasing baryon density and 
in the general case of isospin asymmetric matter shows a splitting 
between $K^{0}$ and $K^{+}$, see Fig. \ref{kmass}.

\begin{figure}[t]
\unitlength1cm
\begin{picture}(8.,7.)
\put(2.0,0.0){\makebox{\epsfig{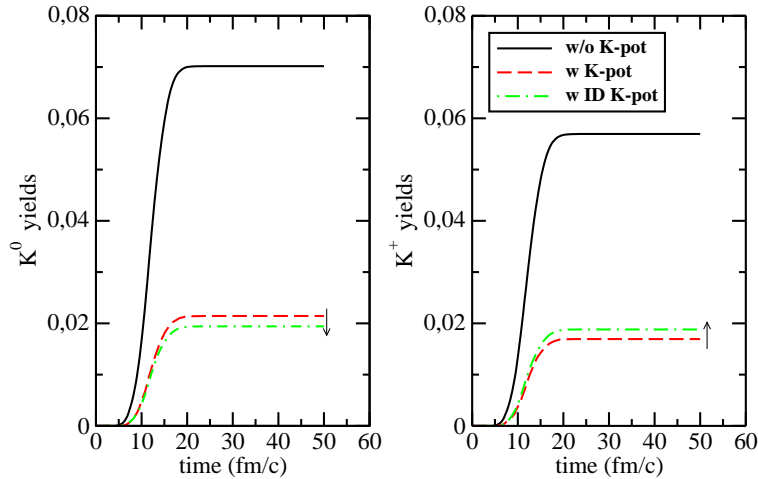}}}
\end{picture}
\caption{Time evolution of the $K^{0}$ (left panel) and $K^{+}$ (right panel)
multiplicities for the same reaction as in Fig.~\protect\ref{deltapions}. 
Calculations without 
(w/o K-pot, solid), with (w K-pot, dashed) and with the isospin dependent 
(w ID K-pot,dotted-dashed) kaon potential are shown. In all the 
cases the free choice for $\sigma_{inel}$ is adopted. 
}
\label{kpot}
\end{figure}

We are presenting here several K-production results in $ab~initio$
collision simulations using the Chiral determination of the K-potentials,
$ChPT$.
Fig. \ref{kpot} shows the time dependence of the two isospin states of the 
kaon with respect to the role of the kaon potential and its isospin 
dependence. First of all, the repulsive kaon potential considerably 
reduces the 
kaon yields, at least in this $ChPT$ evaluation. 

\begin{figure}[t]
\unitlength1cm
\begin{picture}(8.,7.)
\put(2.0,0.0){\makebox{\epsfig{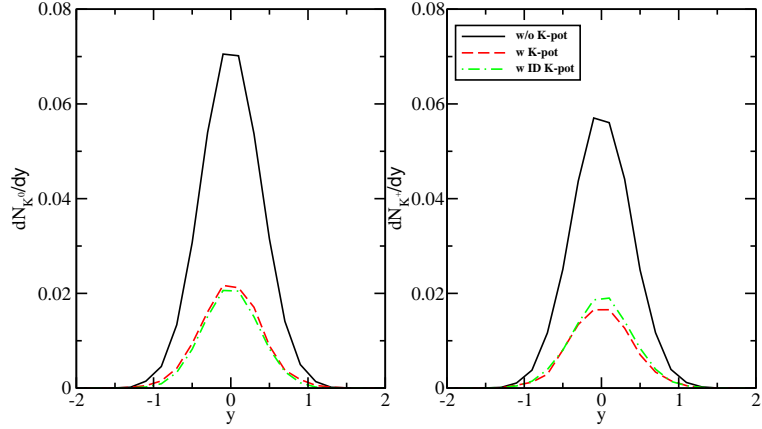}}}
\end{picture}
\caption{Same as in Fig. \protect\ref{kpot}, but for the rapidity 
distributions.
}
\label{kpotrap}
\end{figure}

The inclusion of the isospin dependent part of the kaon potential 
slightly modifies the kaon yields, towards a larger $K^+$
production in neutron-excess matter. However by comparing to the corresponding 
isospin dependence of the in-medium kaon energy, see Fig. \ref{kmass}, 
the effect is less pronounced in the dynamical situation. This is due 
to the fact that in heavy ion collisions the local asymmetry in the 
interacting region
varies with time, see \cite{ferini2}. In particular, it decreases with 
respect to the 
initial asymmetry because of partial isospin 
equilibration due to stopping and inelastic processes with associated 
isospin exchange. This is reflected also in the kaon rapidity 
distributions, see Fig. \ref{kpotrap}, where the role of the kaon 
potential is crucial, but not its isospin dependence.

\begin{figure}[t]
\unitlength1cm
\begin{picture}(8.,7.5)
\put(2.0,0.0){\makebox{\epsfig{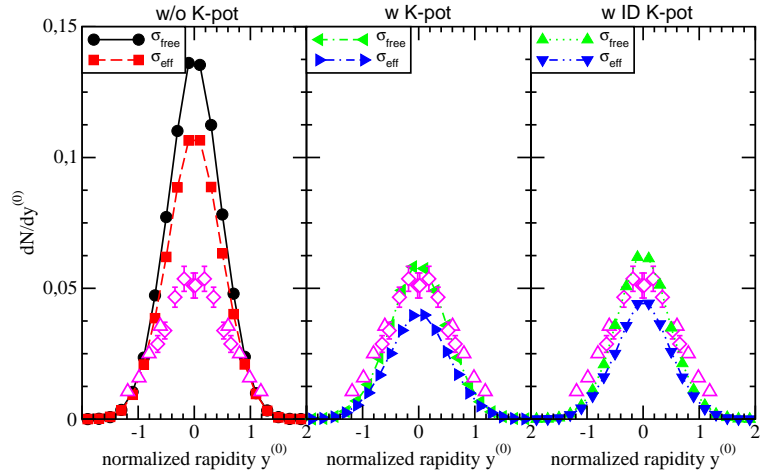}}}
\end{picture}
\caption{ $K^{+}$ rapidity distributions for 
semi-central($b<4~fm$) Ni+Ni reactions 
at 1.93 AGeV. Theoretical calculations (as indicated) are compared with 
the experimental data of FOPI (open triangles) and KaoS (open diamonds) 
collaborations \protect\cite{expf,expk}.
}
\label{kNiNi}
\end{figure}

As we have already seen even in-medium modifications of inelastic cross 
sections
are affecting the kaon absolute yields, so it appears of interest to
look at the combined effects. 
For that purpose we have performed calculations for a semi-central ($b<4 fm$)
Ni+Ni system at $1.93 AGeV$, where data are existing from the FOPI
\cite{expf} and KaoS \cite{expk} collaborations. The results for $K^+$ 
rapidity distributions, 
compared to 
experimental 
data, are shown in Fig.\ref{kNiNi}. 

We observe that although the 
kaon yields
are reduced when using the in-medium inelastic cross section, we are still 
rather far away from the data, left panel of Fig. \ref{kNiNi}. 
We note that the reduction due to the density dependence of the
effective inelastic cross sections is rather moderate here with
respect to that of the heavier Au-system (see Fig. \ref{krap}).
for kaons). This is due to the less compression achieved for
the lighter Ni-systems.
The inclusion of the kaon potential, without (central panel) and with
(right panel) isospin dependence, is further suppressing the $K^+$ yield,
towards a better agreement with data, as expected for the repulsive 
behavior at high density.

In fact the results obtained 
with kaon potentials and effective cross sections seem to underestimate the 
data.
This could be an indication that the $ChPT$ K-potentials are too repulsive
at densities around $2\rho_0$ where kaons are produced, see \cite{ferini2}.
We like to remind that the parameters of $ChPT$ potentials are essentially 
derived from free space considerations. When we follow a more consistent
$RMF$ approach, directly linked to the effective Lagrangians used to describe
bulk properties of the nuclear matter as well as the relativistic transport 
dynamics, we see less repulsion, bottom curves in Fig. \ref{kmass} 
(left panel).
This is valid also for the isospin dependent part of the K-potentials, that
more directly will affect the $K^0/K^+$ yield ratio. We see from the same 
Fig. \ref{kmass} (right panel) that in the $RMF$ frame this splitting 
is reduced to a few
percent for all the different isovector interactions.
The conclusion is that when kaon potentials are evaluated within a consistent
effective field approach we have a better agreement with data for absolute 
yields,
 with a very similar reduction of the $K^0$ and $K^+$ rates. This is important
for the yield ratio, that then should be not much sensitive to the in-medium
effects on kaon propagation.  

\begin{figure}[t]
\unitlength1cm
\begin{picture}(8.,8.5)
\put(2.0,0.0){\makebox{\epsfig{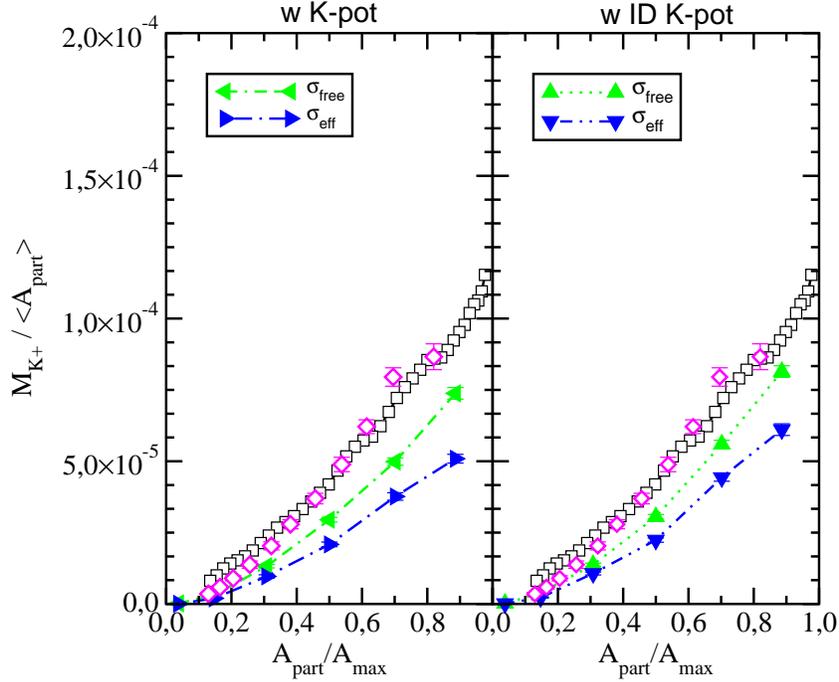}}}
\end{picture}
\caption{$K^{+}$ centrality dependence in Au+Au reactions at 
$1$ AGeV incident energy. 
Our theoretical calculations (as indicated) are compared with 
KaoS data from \protect\cite{expa} (open diamonds) and with results 
of the T\"{u}bingen group (open squares). 
}
\label{kpotcentr}
\end{figure}

  A similar conclusion on K-potential effects, obtained within the
$ChPT$ approach, can be drawn from the 
centrality dependence
of the $K^+$ yields shown in Fig. \ref{kpotcentr} in the case of Au+Au
collisions at 1 AGeV beam energy and compared with KaoS data \cite{expa}.
The trend in centrality can be reproduced by 
all theoretical calculations (with different cross sections), however, 
all of them seem to underestimate the 
experimental yields.

In fact 
we have to mention that another possible source of the discrepancy
with data can be that in all our simulations only the 
lowest mass resonance $\Delta (1232 MeV)$ has been dynamically included. 
Transport calculations from other groups, that take care also of the 
$N^{*}(1440 MeV)$ resonance, are getting an enhancement of the $K^{+}$ 
yield for Au+Au collisions at 1 AGeV incident 
energy \cite{fuchs06}. This significant dependence of the kaon yields on the 
$N^{*}$ resonance comes from the $2$-pionic $N^{*}$-decay channel, 
i.e. $N^{*}\longrightarrow \pi\pi N$. Therefore, since the most 
important channels of kaon production are the pionic ones, we can 
 expect some underestimation of the  absolute yields in our 
calculations. Just to confirm this point, in Fig. \ref{kpotcentr} 
we report also transport 
results from 
the T\"{u}bingen group, in which all resonances are 
accounted for \cite{fuchs06}. We finally remark that the inclusion of other
nucleon resonances in neutron-rich matter will further contribute to
increase the $K^0$ yield through a larger intermediate $\pi-$ production.
This can contribute to compensate the opposite effect of isospin dependent 
part of the K-potentials
on the $K^0/K^+$ yield ratios.

\subsection{Pionic and Strangeness Ratios}

A crucial question is whether particle yield {\it ratios} are
influenced by in-medium effects both on inelastic cross section and 
kaon potentials. 
This point is of major importance particularly for kaons, 
since ratios of particles with strangeness have been widely 
used in determining the nuclear EoS at supra-normal density. Relative 
ratios of 
kaons between
different colliding systems have been utilized in determining the
isoscalar sector of the nuclear EoS \cite{fuchs}. More recently,
the $(\pi^{-}/\pi^{+})$- and $(K^{0}/K^{+})$-ratios have been proposed
in order to explore the high density behavior of the symmetry energy,
i.e. the isovector part of the nuclear mean field
\cite{ferini,ferini2,stoecker,bao}.

\begin{figure}[t]
\unitlength1cm
\begin{picture}(8.,7.)
\put(2.0,0.0){\makebox{\epsfig{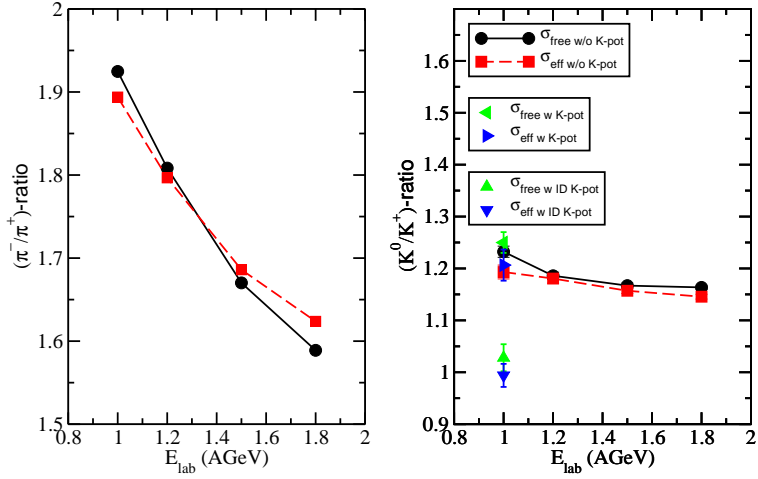}}}
\end{picture}
\caption{Energy dependence of the $\pi^{-}/\pi^{+}$ (left panel) and
$K^{0}/K^{+}$ (right panel) ratios for central ($b=0$ fm) Au+Au 
reactions.
}
\label{ratios}
\end{figure}

Fig. \ref{ratios} shows the incident energy dependence of the pionic 
($\pi^{-}/\pi^{+}$, left panel) and strangeness 
($K^{0}/K^{+}$, right panel) ratios for the different choices of 
inelastic cross sections and kaon potentials, as 
widely discussed in the previous sections. 

First of all, a rapid decrease of the pionic ratio with increasing 
beam energy is observed, related to the opening of secondary rescattering 
processes (reabsorption/recreation of pions with associated isospin 
exchange) channels. 
The corresponding 
strangeness ratio depends only moderately on beam energy due to the 
absence of secondary interactions with the hadronic environment. 

The pionic ratio is partially affected by the in-medium effects 
of $\sigma_{inel}$, as it can be seen in the left panel of Fig.\ref{ratios}. 
Its slope is slightly changing with respect to beam energy. The situation 
is similar for the strangeness ratio, which actually appears even 
more $robust$ vs. in-medium modifications, even with the kaon potentials.
This can be seen in the 
right panel of Fig.\ref{ratios}, where for all the considered beam energies 
the ratio remains almost unchanged. Such a result is consistent with those 
of the 
previous sections, where it was found that the absolute kaon yields decrease 
in the same way when the effective $\sigma_{inel}$ are applied and when
the K-potentials are included. 

The different sensitivity to variations in the inelastic cross sections
of pionic vs. strangeness ratios can be easily understood. For the large
rescattering and lower masses pions can be produced at different times
during the collision, in different density regions. At variance kaons are
mainly produced at early times in a rather well definite compression stage, 
i.e. in a source with a more uniform high density, and so the 
density dependence
of the inelastic cross sections will affect in the same way neutral and 
charged kaon yields, leaving the ratio unchanged.
At this level of 
investigation one could argue that the strangeness ratio is a very 
promising observable in determining the nuclear EoS and particularly 
its isospin dependent part. This has been also the main conclusion in 
Ref. \cite{ferini2}. 

However, a strong isospin dependence of the kaon potentials could
directly affect  the ratio, since the $K^0$ and $K^+$ rates will be 
modified in opposite ways. This is shown by the two triangle points
at 1 AGeV in the right panel of Fig. \ref{ratios}. As already discussed this
large isospin dependence of the kaon potential, clearly present in the $ChPT$
evaluation, is greatly reduced in a consistent mean field approach, see 
Fig. \ref{kmass} and the arguments presented in Section 3.
In any case this point deserves more detailed studies. 
We plan to perform $ab~initio$ kaon-production simulations within the 
$OBE/RMF$ evaluation of kaon potentials, with an isospin part fully consistent
with the isovector fields of the Hadronic Lagrangians used for the
reaction dynamics. 

An interesting final comment is that the sensitivity of the strangeness ratio
to the isovector part of the nuclear EoS remains even when strong isospin
dependence of the kaon potentials is inserted, as in the $ChPT$ case.


\begin{table}[t]
\begin{center}
\begin{tabular}{|l|c|c|c|}
\hline\hline 
                 & $NL$   & $NL\rho$  & $NL\rho\delta$   \\ 
\hline\hline
   $K^{0}/K^{+}$ (w/o K-pot) &     1.24 ($\pm$ 0.02)  & 1.35 ($\pm$ 0.01)  &
                 1.43 ($\pm$ 0.02)  \\ 
\hline\hline
   $K^{0}/K^{+}$ (w ID K-pot) & 1.02 ($\pm$ 0.03) & 1.22 ($\pm$ 0.04)   &
                 1.34 ($\pm$ 0.05)  \\ 
\hline\hline
\end{tabular}
\end{center}
\vskip 0.5cm
\caption{\label{table2} 
Sensitivity of the strangeness ratio $K^{0}/K^{+}$  to the isospin dependent 
kaon potential and to the isovector mean field 
($NL$, no isovector fields, $NL\rho$ and $NL\rho\delta$). The considered 
reaction is a 
central ($b=0$ fm) Au+Au collision at 1 AGeV incident energy.
}
\vskip 0.5cm
\end{table}

 In order to check this, we 
have repeated for Au+Au at 1 AGeV incident energy the calculations 
by varying the isovector part of the nuclear mean field. As in 
Refs. \cite{ferini,ferini2}, three options for the isovector mean field 
have been applied: the $NL$ (no isovector fields), $NL\rho$ and 
$NL\rho\delta$ parametrizations, 
but now including the isospin effect in the kaon potential in the
$ChPT$ evaluation. The options 
of the symmetry energy differ from each other in the high density 
stiffness. $NL$ gives a relatively soft $E_{sym}$, $NL\rho\delta$ a 
relatively stiff one, and $NL\rho$ lies in the middle between the other 
limiting cases \cite{ferini}. Table \ref{table2} shows the strangeness 
ratio as function of these different cases for the isovector mean field, 
keeping now constant the other parameters (free $\sigma_{inel}$, isospin 
dependent kaon potential). The ratio, indeed, strongly decreases when 
the isospin part in the kaon potential is accounted for. The more interesting 
result is, however, that the relative difference between the different 
choices of the symmetry energy remains stable. This can be understood from 
the fact that in the kaon self energies the isospin sector 
contains only the isospin densities and currents without additional parameters 
such meson-nucleon coupling constants. Since the local asymmetry does not 
strongly vary from one case to the other 
($NL$, $NL\rho$, $NL\rho\delta$), one would expect a robustness of the EoS 
dependence. 
Thus we conclude that the strangeness ratio appears to be well suited in 
determining 
the isovector EoS, however, a fully consistent mean field approach
is still missing.

\section{Conclusions}

We have investigated the role of the in-medium modifications of
the inelastic cross section and  of the 
kaon mean field potentials on particle production in intermediate
energy heavy ion collisions within a covariant transport equation
of a Boltzmann type. We have used for both, the elastic and
inelastic NN cross sections the same DBHF approach which provide
in a parameter free manner the in-medium modifications of the
imaginary part of the self energy in nuclear matter. The kaon 
potential has been evaluated in two ways, following a Chiral Perturbative
approach and an Effective Field scheme, considering valence quark-meson
couplings. We have
applied these modifications of the cross sections and kaon potentials 
to the collision integral of the transport equation
and analyzed Au+Au and Ni+Ni collisions at intermediate relativistic
energies around the kaon threshold energy.

Our studies have shown a good sensitivity of the particle
multiplicities and rapidity distributions of pions and kaons. In
particular, a moderate reduction for pions has been
seen when the in-medium effects in the inelastic cross section are
accounted for. The pion yields are still overestimating the inclusive data
while we have a very nice agreement with the pion spectra and
multiplicities at mid-rapidity.
The latter point is important for trusting the kaon production, mainly due 
to secondary pion collisions at mid-rapidity.

At variance the kaon ($K^{0,+}$) yields show a larger sensitivity
to the reduction of the inelastic cross sections, with a
decrease of about 30 \%. However we see that the introduction of a repulsive
kaon potential is essential in order to reproduce even inclusive data.

We have then focused our attention on $\pi^- / \pi^+$ and $K^0/K^+$ yield 
ratios, recently suggested as good probes of the isovector part of the
$EoS$ at high densities. The
pionic ratios, due to their strong secondary interaction processes
with the hadronic environment, show a  dependence on the
density behavior of the inelastic cross sections. A further selection of the
production source, i.e. a transverse momentum discrimination, could be required
in order to have a more reliable probe of the nuclear $EoS$.

The situation appears more favorable for the kaon ratios. 
In fact we find that the multiplicities of $K^{0}$ and $K^{+}$ are
influenced in such a way that their ratio is not affected by  the
density dependence of the inelastic cross sections. This is due to
the long mean free path of the $K^{0,+}$ that are produced only in
the compression stage of the collision \cite{ferini2}.
The effects of the in medium kaon potentials are also largely compensating
in the $K^0/K^+$ yield ratio, due to the similar repulsive field seen
by $K^{0}$ and $K^{+}$ mesons.
Such a result can be modified by the isospin dependence of the kaon potentials
which is expected to act in opposite directions for neutral and charged
kaons rates. Actually this is a rather stimulating open problem. In our 
analysis
with two completely different approaches, $ChPT$ vs. $RMF$, we get a good 
agreement for the isoscalar kaon potential but a rather different prediction
for the isovector part.

 However, a study in terms 
of the different choices of the isovector kaon field has shown that the 
relative dependence of the strangeness ratios on the stiffness of the 
isovector nuclear EoS remains 
a well robust observable. This is an important issue in determining 
the high density behavior of the symmetry energy in more systematic 
analyses in the future, when more experimental data will be available.\\

{\it Acknowledgments.} This work is supported by BMBF, grant 06LM189
and the State Scholarships Foundation (I.K.Y.). It is also co-funded by 
European Union Social 
Fund and National funded Pythagoras II - EPEAEK II, under project 80861.
 One of the authors (V.P.) would like to thank
H.H. Wolter and M. Di Toro for the warm hospitality during her
short stays at their institutes.


\end{document}